\documentclass[twocolumn,showpacs,preprintnumbers,amsmath,amssymb,pra]{revtex4-1}
\usepackage[dvips]{graphics,graphicx,epstopdf}
\usepackage{mathrsfs}
\usepackage{pdfpages}
\usepackage{float}
\usepackage{braket}
\usepackage{natbib}
\begin{document}
\title{General ultracold scattering formalism with isotropic spin orbit coupling}
\author{Su-Ju Wang}
\email{wang552@purdue.edu}
\author{Chris H. Greene}
\email{chgreene@purdue.edu}
\affiliation{Department of Physics and Astronomy, Purdue University, West Lafayette, IN
47907, USA}
\date{\today}

\begin{abstract}
A general treatment of ultracold two-body scattering in the presence of isotropic spin-orbit coupling (SOC) is presented. Owing to the mixing of different partial wave channels, scattering with SOC is in general a coupled multichannel problem. A systematic method is introduced to analytically solve a class of coupled differential equations by recasting the coupled channel problem as a simple eigenvalue problem. The exact Green's matrix in the presence of SOC is found, which readily gives the scattering solutions for any two identical particles in any total angular momentum subspace having negligible center of mass momentum. Application of this formalism to two spin-1 bosons shows the ubiquitous low energy threshold behavior for systems with isotropic SOC. A modified threshold behavior shows up, which does not occur for the spin-orbit coupled spin-1/2 system. We also confirm the parity-breaking mechanism for the spontaneous emergency of handedness, that has been proposed by Duan {\em et~al.} \cite{Duan}. Additionally, a two-body bound state is found for any arbitrarily small and negative scattering length. Our study sheds light on the few-body side of SOC physics and provides one step towards understanding ultracold scattering in a non-Abelian gauge field.
\end{abstract}


\maketitle

\section{Introduction}
Since the first discoveries of Bose-Einstein condensates and degenerate Fermi gases \cite{BEC1,BEC2}, ultracold atomic systems have emerged as a new class of highly-controllable systems that can serve as quantum simulators of traditional condensed matter systems. Spin-orbit coupling in particular is an important ingredient in topological insulators as well as many other intriguing phenomena (see reviews \cite{toI,toII}). Therefore, the development of a way to create {\em synthetic} gauge fields in {\em neutral} cold atom systems has been a fundamental advance in recent years \cite{Dalibard,gf2}. In 2011, the Spielman group at NIST successfully engineered equal Rashba and Dresselhaus spin-orbit coupling in a Bose-Einstein condensate by dressing two of the hyperfine spin states of $^{87}$Rb with two counter-propagating laser fields \cite{YJLin}. This achievement has created a new paradigmatic system and a new tool for manipulating an ultracold quantum gas, and has provided a new direction for the study of many-body and few-body systems \cite{socrev,fbmb}.

Spin-orbit coupling is characterized by its unusual energy dispersion relations. Early on, the double minimum energy dispersion was proposed to generate macroscopic quantum superposition states with repulsive interatomic interactions \cite{Higbie}. The non-quadratic energy dispersion relation modifies the density of states, and has been shown to significantly change the bound state spectrum \cite{Cuimixed,bound,ZhangZhang}. For example, it has been proved theoretically that bound states for two spin-1/2 fermions exist for an arbitrarily weak attraction in the presence of 3D isotropic spin orbit coupling \cite{bound}. Also the energy spectrum of a harmonically trapped two-atom system is studied \cite{conf1,Blume,Blume2}. Not only the two-body bound state spectrum, but also the scattering formalism becomes modified since SOC exists to infinite distance and this will be the main topic of our study here.

Inspired by the recent work done by Duan {\em et al.} \cite{Duan}, we generalize their treatment of the scattering of two spin-1/2 fermions in the presence of isotropic SOC.  Our formulation applies to any two identical bosons or fermions of arbitrary spin, for arbitrary values of the total angular momentum of the system. An advantage to the choice of isotropic SOC, which is a 3D analog of Rashba SOC, is that it has higher symmetry than other types of SOC, and is more closely related to the cases in condensed matter physics \cite{Weyl1,Weyl2}. The conservation of total angular momentum allows us to develop a fully analytical treatment of scattering theory in the presence of SOC. The generalization of two-body scattering to higher spin atoms can extend our understanding to higher spin physics having no counterparts in condensed matter systems. For instance, a system of spin-3/2 fermions with contact potential interactions has been shown to have exact SO(5) symmetry, and a novel quartetting order (a four-fermion version of Cooper pairing) has been proposed \cite{Wu}. Understanding the two-body physics also paves the way to more interesting varieties of universal Efimov physics \cite{Borromean,Zhaitrimer,3bnew}. Although Rashba-type SOC has not yet been realized experimentally, proposals have been made that are based on adding more laser fields \cite{pro_1} or else by applying magnetic pulses \cite{Ueda,pro_2} to imprint an engineered phase onto the atoms. 

This paper is organized as follows: Sec.~II presents a systematic way to formulate the multichannel 2-body scattering problem with SOC present, and outlines the route to extract the scattering information. The first step is a derivation of an analytical expression for the free Green's matrix with SOC. When the atoms interact through a regularized $s$-wave interaction, which is an excellent assumption in the ultracold regime, the Lippmann-Schwinger equation can then be cast into a simple form having a closed form solution. Utilization of the Green's matrix and the Lippmann-Schwinger equation enables the analytical scattering wave functions to be found, and the scattering properties extracted. Sec.~III applies our methodology to a system of two identical spin-1 bosonic atoms, and derives the scattering cross sections. An unusual type of threshold behavior is seen to emerge in the low energy scattering cross section. Sec.~III confirms the spontaneous emergence of handedness in this type of system having no parity symmetry. Discussion about two-body bound states is included too. Finally, Sec.~IV discusses our conclusions.
%%%%%%%%%%%%%%%%%%%%%%%%%%%%%%%%%%%%%%%%%%%%%%%%%%%%%%%%%%%%%%%%%%%%%%%%
%Model
%%%%%%%%%%%%%%%%%%%%%%%%%%%%%%%%%%%%%%%%%%%%%%%%%%%%%%%%%%%%%%%%%%%%%%%%
\section{Model}
For identical particles interacting with each other in the presence of isotropic 3D spin-orbit coupling, the two-body Hamiltonian is expressed as
\begin{equation}
H_{\text{2b}}=\frac{\hbar^2\vec{k}^2_1}{2m}+\frac{\hbar^2\lambda}{m}\vec{k}_1\cdot{\vec{s}_1}+\frac{\hbar^2\vec{k}^2_2}{2m}+\frac{\hbar^2\lambda}{m}\vec{k}_2\cdot{\vec{s}_2}+V(\vec{r}_1-\vec{r}_2),
\end{equation}
where $m$ is the atomic mass, $\lambda$ is the strength of the spin-orbit coupling and $V(\vec{r}_1-\vec{r}_2)$ is the interatomic interaction. The operator $\vec{s}_1$ and $\vec{s}_2$ are the hyperfine spin operators for atom $1$ and atom $2$; hereafter these are referred to simply as spin. Since the total momentum in the system is conserved, the center of mass motion and the relative motion can be decoupled. The two-body Hamiltonian can be rewritten as usual using the center of mass momentum operator $\vec{P}=\vec{p}_1+\vec{p}_2$, and the relative momentum operator $\vec{p}=(\vec{p}_1-\vec{p}_2)/2$.  The two-body Hamiltonian then becomes
\begin{align}
\nonumber
H_{\text{2b}}=&H_{\text{com}}+H_{\text{rel}}=\frac{\vec{P}^2}{4m}+\frac{\hbar\lambda}{2m}\vec{P}\cdot (\vec{s}_1+\vec{s}_2)\\
&+\frac{\vec{p}^2}{m}+\frac{\hbar\lambda}{m}\vec{p}\cdot (\vec{s}_1-\vec{s}_2)+V(\vec{r}_1-\vec{r}_2).
\end{align}
Although the center of mass momentum and the relative motion can be separated out, the relative motion is generally coupled to the center of mass motion via the spin degrees of freedom. To simplify the present calculation, the remainder of this paper is formulated within the center of mass frame and we focus on the case of $\vec{P}=0$. (Note also that the orbital angular momentum of center of mass is $L_{\vec{R}}=0$); thus, $H_{\text{2b}}=H_{\text{rel}}+V(\vec{r}_1-\vec{r}_2)$. When the  center of mass momentum is nonzero, this breaks the continuous rotational invariance of relative energy spectra and degeneracies of relative band energies are lifted, although we do not discuss it here in detail. 

A key first step is to solve the relative Schr\"{o}dinger equation in the absence of interactions. Since the relative momentum commutes with the non-interacting Hamiltonian, it is advantageous to solve it in momentum space and then Fourier transform the solution back to position space. Taking spin-1 bosons as an example, the non-interacting two-body states are:
\begin{align}
\nonumber
&\braket{\vec{r}|\zeta,\xi;\vec{k}}=\\
\label{planewavesol}
&\;\;\;\;\;\;\;\;\;\;\;\;\;\frac{1}{\sqrt{2}}\big(\ket{\zeta,\hat{k}}_1\ket{\xi,-\hat{k}}_2 e^{i\vec{k}\cdot \vec{r}}+\ket{\xi,-\hat{k}}_1\ket{\zeta,\hat{k}}_2 e^{-i\vec{k}\cdot \vec{r}}\big),
\end{align}
where $\ket{\zeta,\hat{k}}$ and $\ket{\xi,\hat{k}}$ are one of the following single-particle states:
\begin{align}
\label{hn}
&\ket{-,\hat{k}}=\begin{pmatrix}  e^{-i\phi_{\vec{k}}}\frac{(1-\cos\theta_{\vec{k}})}{2} \\  -\frac{\sin\theta_{\vec{k}}}{\sqrt{2}} \\  e^{i\phi_{\vec{k}}}\frac{(1+\cos\theta_{\vec{k}})}{2}  \end{pmatrix},\;E_-=\frac{\hbar^2k^2}{2m}-\frac{\hbar^2\lambda k}{m}\\
\label{h0}
&\ket{0,\hat{k}}=\begin{pmatrix}  -e^{-i\phi_{\vec{k}}}\frac{\sin\theta_{\vec{k}}}{\sqrt{2}} \\  \cos\theta_{\vec{k}} \\  e^{i\phi_{\vec{k}}}\frac{\sin\theta_{\vec{k}}}{\sqrt{2}}  \end{pmatrix},\;E_0=\frac{\hbar^2k^2}{2m}\\
\label{hp}
&\ket{+,\hat{k}}=\begin{pmatrix}  e^{-i\phi_{\vec{k}}}\frac{(1+\cos\theta_{\vec{k}})}{2} \\  \frac{\sin\theta_{\vec{k}}}{\sqrt{2}} \\  e^{i\phi_{\vec{k}}}\frac{(1-\cos\theta_{\vec{k}})}{2}  \end{pmatrix},\;E_+=\frac{\hbar^2k^2}{2m}+\frac{\hbar^2\lambda k}{m}
\end{align}
where $\theta_{\vec{k}}$ and $\phi_{\vec{k}}$ describe the direction of the particle's motion along $\hat{k}$. The eigenstates are expressed in the basis of $\{\ket{1,1},\ket{1,0},\ket{1,-1}\}$, which are the eigenstates of the ${s}_z$ operator for each atom. The three states in Eq.(\ref{hn})$\sim$(\ref{hp}) are also eigenstates of the helicity operator, $h=\vec{p}\cdot \vec{s}/p$, with eigenvalues  -1, 0, and 1. In general, the eigenvalues range from $-s$, $-s+1$, ... to $s$ for spin $\vec{s}$. The helicity states can be pictured in an intuitive way as follows: when a spin, $\vec{s}$, moves along direction $\hat{k}$, there are $(2s+1)$ possible spin configurations. The maximum (minimum) helicity state represents the state when the particle's spin is in parallel (antiparallel) to the direction of its motion. For the same canonical momentum, when spin is aligned with its momentum, the state has the highest eigenvalue. In the article, we will mainly discuss how the particles with definite helicity are going to be scattered to different helicity states through helicity non-conserving interaction.  
\begin{figure}
\begin{center}
\includegraphics[width=2.7in]{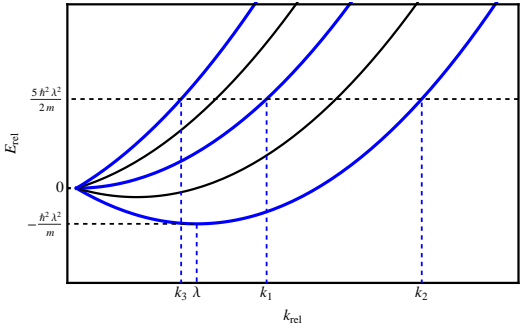}
\caption{(Color online) The energy-momentum dispersion relation for the spin-1 bosons is depicted. There are 9 bands in total. However, bands with the same resultant spin component along the direction of the relative motion are degenerate. The degeneracies from left to right are 1, 2, 3, 2, and 1. In $\ket{J=0}$ subspace, only the thick blue bands are involved.}
\label{enerydis}
\end{center}
\end{figure}

The methodology to solve the two-body scattering problem is sketched below: we first calculate the regular and irregular solutions of the non-interacting system that satisfy the correct boundary conditions, and then use those solutions to construct the free-particle Green's matrix with isotropic spin-orbit coupling. The Green's matrix is then used in the Lippmann-Schwinger equation to solve for the scattered wave functions. 
%All the desired scattering information can be extracted from there.
%%%%%%%%%%%%%%%%%%%%%%%%%%%%%%%%%%%%%%%%%%%%%%%%%%%%%%%%%%%%%%%%%%%%%%%%
%Green's function with SOC
%%%%%%%%%%%%%%%%%%%%%%%%%%%%%%%%%%%%%%%%%%%%%%%%%%%%%%%%%%%%%%%%%%%%%%%%
\subsection{Green's matrix with spin-orbit coupling}
The crucial symmetry in this isotropic spin-orbit system is the conservation of total angular momentum. This allows us to expand the solutions in a complete basis set having a fixed value of the total angular momentum quantum number, $J$. Because both the orbital angular momentum and the spin angular momentum would be conserved in the absence of spin-orbit coupling, tensor spherical harmonics \cite{VMK} are adopted as the basis set. These are simultaneous eigenstates of $\{\vec{J}^2, {J}_z,\vec{L}^2,\vec{S}^2\}$, where $\vec{L}$ is the (relative) orbital angular momentum and $\vec{S}=\vec{s}_1+\vec{s}_2$ is the total spin angular momentum. In terms of this basis set, spin-orbit coupling simply mixes states with different $\{L,S\}$-values, which label these basis functions. Consequently the Hamiltonian matrix elements in this basis set have nonzero off-diagonal elements. The tensor spherical harmonics are defined as
\begin{equation}
Y_{JM}^{LS}(\theta,\phi)=\sum_{m_L,m_S}C^{JM}_{Lm_L,Sm_S} Y_{Lm_L}(\theta,\phi)\chi(S,m_S)
\end{equation}
where $C^{JM}_{Lm_L,Sm_S}$ is the Clebsch-Gordan coefficient, $Y_{Lm_L}(\theta,\phi)$ is the spherical harmonics, and $\chi(S,m_S)$ is the spin state for total spin $S$. 
Any wave function can be expanded in this tensor spherical harmonics basis set, 
\begin{equation}
\Psi^{JM}_\eta(r,\theta,\phi)=
\sum_{\{L,S\}} \frac{u^{JM}_{\{L,S\},\eta}(r)}{r}\times Y_{JM}^{LS}(\theta,\phi),
\end{equation}
where $u(r)$ is the reduced radial wave function and the index $\eta$ represents different independent solutions. The matrix element of the kinetic energy operator is easily evaluated, and the result is familiar:
\begin{align}
\nonumber
&\langle(L',S')J'M' |\frac{\vec{p}^2}{m}|(L,S)JM\rangle=\\
&(-\frac{\hbar^2}{m}\frac{d^2}{dr^2}+\frac{L(L+1)\hbar^2}{mr^2})\delta_{J,J'}\delta_{M,M'}\delta_{L,L'}\delta_{S,S'}.
\end{align}
Matrix elements of the spin-orbit coupling term are evaluated using the Wigner-Eckart theorem in the convention of Ref.~\cite{VMK}: 
\begin{align}
\nonumber
&\langle (L',S')J'M'|\vec{p}\cdot(\vec{s}_1-\vec{s}_2)|(L,S)JM\rangle= (-1)^{J+L+S'}\times\\
\label{eqWE}
&\;\;\;\; \delta_{JJ'}\delta_{MM'} \langle L'\|{p^{(1)}}\|L\rangle  \langle S'\|s_1^{(1)}-s_2^{(1)}\|S\rangle 
\begin{Bmatrix}
L'&L&1    \\
S&S'&J   \end{Bmatrix},
\end{align}
where the curly bracket denotes the $6j$ symbol and the double bars stand for reduced matrix elements, which are defined by
\begin{align}
\langle L'\|{p}^{(1)}\|L\rangle=\frac{\langle L'm_L'|{p}^{(1)}_q|Lm_L\rangle}{C^{L'm_L'}_{Lm_L,1q}}\sqrt{2L'+1},\\
\langle S'\|{s^{(1)}}\|S\rangle=\frac{\langle S'm_S'|{s}^{(1)}_q|Sm_S\rangle}{C^{S'm_S'}_{Sm_S,1q}}\sqrt{2S'+1},
\end{align}
where the superscript inside the parentheses is the rank of the operator and the subscript means the $q$th component of that tensor operator. 
All the dependence on magnetic quantum numbers occur now in the Clebsch-Gordan coefficients, in the usual spirit of the Wigner-Eckart theorem. Application of some straightforward angular momentum algebra yields the matrix element of $\vec{p}\cdot (\vec{s}_1-\vec{s}_2)$,
\begin{widetext}
\begin{align}
\nonumber
&\langle (L',S')J'M'|\vec{p}\cdot (\vec{s}_1-\vec{s}_2)|(L,S)JM\rangle=\delta_{JJ'}\delta_{MM'}  \sqrt{(2S+1)(2S'+1)} \begin{Bmatrix}
L'&L&1\\
S&S'&J   \end{Bmatrix}(-1)^{J+L+S'+s_1+s_2}
\\
\nonumber
&\times \bigg[
-(-1)^{S}
\sqrt{s_1(s_1+1)(2s_1+1)}\times
\begin{Bmatrix}
s_1&s_2&S\\
S'&1&s_1   \end{Bmatrix}
+(-1)^{S'} \sqrt{s_2(s_2+1)(2s_2+1)}\times
\begin{Bmatrix}
S&S'&1\\
s_2&s_2&s_1 \end{Bmatrix}
\bigg]\\
&\times \left\{ 
  \begin{array}{l l}
    -i\hbar (\frac{d}{d r}-\frac{L}{r})\sqrt{(L+1)} & \quad \text{if $L'=L+1$}\\
   i\hbar (\frac{d}{d r}+\frac{L+1}{r})\sqrt{L} & \quad \text{if $L'=L-1$}
  \end{array} \right.
  \end{align}
 \end{widetext}
The spin-orbit interaction couples states with orbital angular momentum differing by one, which reflects the fact that the rank of the momentum operator is one.
The above matrix elements enable the $n$-coupled radial differential equations to be written for any two spins with any total angular momentum $J$ in their center of mass frame. The number $n$ represents the total number of basis functions in $\ket{J}$ subspace. To solve the coupled differential equations, we make an {\em ansatz} that the regular solutions take the form,
\begin{equation}
\label{reg}
\underline{f}_\eta(r)=\begin{pmatrix}
      c_1k_\eta r j_{L_1}(k_\eta r)\\c_2 k_\eta r j_{L_2}(k_\eta r)\\c_3 k_\eta r j_{L_3}(k_\eta r)\\ \vdots        
\end{pmatrix},
\end{equation}
where $j_{L_i}(k_\eta r)$ is the spherical Bessel function, $L_1$, $L_2$ and so on are the allowed $L$ values from the basis functions of $n=1, n=2, ...$, and $k_\eta$ is the canonical momentum for the $\eta$th independent solution at a fixed incident energy, $E$. The total number of the independent solutions, $\eta$, is equal to the total number of the basis functions, $n$. For non-zero $J$, degeneracies of bands become important and the total number of {\em different} $k_\eta$ may be less than the total number of basis functions concerned. However, this does not affect the form of solution given in Eq.~(\ref{reg}).

Plugging in this ansatz into the coupled differential equations, the differential equations reduce to an eigenvalue problem, $\tilde{\underline{H}}\;\tilde{\underline{\Psi}}=\tilde{E}\tilde{\underline{\Psi}}$, where $\tilde{\underline{H}}$ is given by
\begin{align}
\small
\nonumber
&\langle L',S'|\tilde{H}|L,S \rangle=(-1)^{J+L+S'+s_1+s_2}\sqrt{(2S+1)(2S'+1)}\\
\nonumber
&\;\times \bigg[-(-1)^{S}
\begin{Bmatrix}
s_1&s_2&S\\\nonumber
S'&1&s_1   \end{Bmatrix}\times
\sqrt{s_1(s_1+1)(2s_1+1)}\\\nonumber
&\;+(-1)^{S'}\begin{Bmatrix}
S&S'&1\\
s_2&s_2&s_1 \end{Bmatrix}\times
\sqrt{s_2(s_2+1)(2s_2+1)}\bigg]\times \frac{i \hbar^2\lambda k}{m}
\\
&\;\times
\begin{Bmatrix}
L'&L&1\\
S&S'&J   \end{Bmatrix}
\left\{ 
\begin{array}{l l}
\sqrt{(L+1)} & \quad \text{if $L'=L+1$}\\
\sqrt{L} & \quad \text{if $L'=L-1$},
  \end{array} \right.
\end{align}
and
\begin{equation}
\tilde{\underline{\Psi}}=\{c_1,c_2,c_3,...\}^T.
\end{equation}
The eigenvalues of the matrix $\tilde{H}$ will solve for canonical momenta for fixed energy $E=\tilde{E}+\hbar^2 k_{\eta}^2/m$.
With the standard technique of diagonalization, the solutions of $\{c_1,c_2,\dots\}$ can be found, so are the regular solutions. The solutions irregular at the origin are obtained by replacing the spherical Bessel functions by spherical Neumann functions, $y_{L_i}(k_\eta r)$,
\begin{equation}
\label{irr}
\underline{g}_\eta(r)=\begin{pmatrix}
      c_1 k_\eta r y_{L_1}(k_\eta r)\\c_2 k_\eta r y_{L_2}(k_\eta r)\\c_3k_\eta r y_{L_3}(k_\eta r)\\ \vdots        
\end{pmatrix}.
\end{equation}
The above solutions in Eq.~(\ref{reg}) and Eq.~(\ref{irr}) will be properly energy normalized for an appropriate choice of the momentum-dependent constants, as is carried out below. The reduced radial Green's matrix is shown in appendix A to be
\begin{eqnarray}
\label{eqGred}
{\underline{\mathscr{G}}}(r,r')=\begin{cases}
   \pi \underline{f}(r)\underline{g}^\dagger(r')  & \text{for\,\,\,\,\,} r<r' , \\
  \pi   \underline{g}(r)\underline{f}^\dagger(r') & \text{for\,\,\,\,\,} r>r'.
\end{cases}
\end{eqnarray}
The factor $\pi$ appears because of our choice of normalization. 
More details about energy normalization are also given in appendix A.
%%%%%%%%%%%%%%%%%%%%%%%%%%%%%%%%%%%%%%%%%%%%%%%%%%%%%%%%%%%%%%%%%%%%%%%%
%LS equation 
%%%%%%%%%%%%%%%%%%%%%%%%%%%%%%%%%%%%%%%%%%%%%%%%%%%%%%%%%%%%%%%%%%%%%%%%
\subsection{Lippmann-Schwinger equation}
To solve the scattering wave function for two atoms with isotropic spin-orbit coupling, we apply the Lippmann-Schwinger equation, which is the integral form of the Schr\"odinger equation.
\begin{equation}
{\Psi}(\vec{r})={\Psi}_0(\vec{r})+\int {G}(\vec{r},\vec{r}'){V}(\vec{r}'){\Psi}(\vec{r}')d\vec{r}',
\end{equation}
where $\Psi_0(\vec{r})$ is the non-interacting solution, ${G}(\vec{r},\vec{r}')$ is the free Green's function without 2-body interaction, $V(\vec{r}')$. To compute the wave function that describes scattering processes, we must in general solve the 3-dimensional integral equation in a self-consistent way, which for an arbitrary two-body potential relies on numerics. However, for low energy scattering, the interatomic interaction is well described by the regularized s-wave Fermi pseudo potential, $V(\vec{r})=\frac{4\pi\hbar^2a_s}{m}\delta(\vec{r})\frac{\partial}{\partial r}(r)$, where $a_s$ is the $s$-wave scattering length. It can be shown that the 3D integral equation can be reduced to a 1D radial integral equation, and the scattered wave functions can be obtained in a closed form solution,
\begin{equation}
\label{eqLSradial}
{R}(r)={R}_0(r)+\int_0^\infty {G}(r,r'){V}(r'){R}(r')r'^2dr'.
\end{equation}
Here ${R}_0(r)$ is the free radial two-body wave function. To better illustrate the idea, consider the case of {\em zero} total angular momentum, since in this subspace, the $s$-wave channel is always present.

For any two identical particles with spins having zero total angular momentum, the channel structure is $\{L,S\}=\{0,0\}$, $\{1,1\}$, $\{2,2\}$, $\dots$, and $\{2s_1,2s_1\}$ since from spin statistics $L+S$ has be to even to incorporate the symmetry of identical bosons or fermions. There are $(2s_1+1)$ channels in total. The regularized $s$-wave contact potential is
\begin{align}
\underline{V}(r)=
\begin{pmatrix}
g\frac{\delta(r)}{4\pi r^2}\frac{\partial}{\partial r}r&0&\dots\\
0&0&\dots\\
\vdots&\vdots&\ddots
\end{pmatrix}_{(2s_1+1)\times(2s_1+1)},
\end{align}
where $g=4\pi\hbar^2a_s/{m}=4\pi \tilde{g}$. After applying the operation, $\int_0^\infty dr \tilde{g} \delta{(r)}\frac{\partial}{\partial r}(r)$, to both sides of Eq. (\ref{eqLSradial}), the scattering solutions have the following closed form representation:
\begin{align}
\begin{pmatrix}
R_{1\eta}(r)\\
R_{2\eta}(r)\\
R_{3\eta}(r)\\
\vdots
\end{pmatrix}=
\begin{pmatrix}
R_{0,1\eta}(r)+G_{11}(r,0)\frac{\tilde{g}}{1-\tilde{g} G^\text{reg}_{11}(0,0)}R^\text{reg}_{0,1\eta}(0)\\
R_{0,2\eta}(r)+G_{21}(r,0)\frac{\tilde{g}}{1-\tilde{g} G^\text{reg}_{11}(0,0)}R^\text{reg}_{0,1\eta}(0)\\
R_{0,3\eta}(r)+G_{31}(r,0)\frac{\tilde{g}}{1-\tilde{g} G^\text{reg}_{11}(0,0)}R^\text{reg}_{0,1\eta}(0)\\
\vdots
\end{pmatrix},
\end{align}
where $\eta(=1,2,3,\dots,2s_1+1)$ labels solutions with different canonical momenta, regularized functions $f^{\text{reg}}(0)\equiv \frac{\partial}{\partial r}(rf(r))|_{r\rightarrow0}$, and $f^{\text{reg}}(0,0)\equiv\frac{\partial}{\partial r}(rf(r,0))|_{r\rightarrow0}$.

For systems with nonzero total angular momentum, the algebra can become slightly more involved. The complexity mainly comes from the fact that there are more than one basis function with the same orbital angular momentum but different total spin angular momentum. Degeneracies appear for the two-particle states within some non-zero total angular momentum subspace. This is expected as was already seen in the discussion of sec.~II. Nevertheless, even in this situation, the same methodology can be applied to reduce the coupled differential equations to an eigenvalue problem.
%%%%%%%%%%%%%%%%%%%%%%%%%%%%%%%%%%%%%%%%%%%%%%%%%%%%%%%%%%%%%%%%%%%%%%%%
% spin-1 example
%%%%%%%%%%%%%%%%%%%%%%%%%%%%%%%%%%%%%%%%%%%%%%%%%%%%%%%%%%%%%%%%%%%%%%%%
\section{An example: two spin-1 bosons}
The formalism presented above has been verified to reproduce the results presented by  Duan {\em et~al.}  for two identical spin-1/2 fermions.  The following applies our methodology to the system of two identical spin-1 bosons as a concrete example. One thing worth pointing out is that the normalization factors of the regular/irregular solutions were not written out explicitly in the Duan {\em et~al.} study,  presumably because the factors could be taken to be identical for all the independent solutions. But in the present generalized treatment, it is necessary to keep track of them to ensure flux conservation. 

For two spin-1 bosons with $J=0$, there are only three relevant channels with $\{L,S\}=\{0,0\}, \{1,1\}$ and $\{2,2\}$. The coupled reduced radial differential equations are 
\begin{align}
\small
\nonumber
&\frac{\hbar^2}{m}
\begin{pmatrix}
-\frac{d^2}{dr^2}  &\frac{i2\sqrt{2} \lambda}{\sqrt{3}}(\frac{d}{dr}+\frac{1}{r}) &0 \\
  \frac{i2\sqrt{2} \lambda}{\sqrt{3}}(\frac{d}{dr}-\frac{1}{r})    &  -\frac{d^2}{dr^2}+\frac{2}{r^2}&  \frac{i2 \lambda}{\sqrt{3}}(\frac{d}{dr}+\frac{2}{r})\\
 0&\frac{i2 \lambda}{\sqrt{3}}(\frac{d}{dr}-\frac{2}{r})   &-\frac{d^2}{dr^2}+\frac{6}{r^2} \\
\end{pmatrix}\begin{pmatrix}u^{00}_{00}\\u^{00}_{11}\\u^{00}_{22} \end{pmatrix}\\
\label{eqDE}
&=E \begin{pmatrix}u^{00}_{00}(r)\\u^{00}_{11}(r)\\u^{00}_{22}(r) \end{pmatrix} .
\end{align}
The tridiagonal structure signatures the existence of the spin-orbit coupling. 
Assuming the regular solution has this form,
$\{u^{00}_{00}(r),u^{00}_{11}(r),u^{00}_{22}(r)\}^T=\{c_1 k r j_{0}(kr),c_2k rj_1(kr),c_3krj_{2}(kr) \}^T,$
the following eigenvalue problem is obtained.
\begin{align}
\label{eqevp}
\begin{pmatrix}
\frac{\hbar^2k^2}{m}-E&  2i \hbar^2\sqrt{\frac{2}{3}}\frac{\lambda k}{m} &0 \\
  -2i \hbar^2\sqrt{\frac{2}{3}}\frac{\lambda k}{m}   &  \frac{\hbar^2k^2}{m}-E& 2i \hbar^2\sqrt{\frac{1}{3}}\frac{\lambda k}{m}\\
 0& -2i \hbar^2\sqrt{\frac{1}{3}}\frac{\lambda k}{m}   &\frac{\hbar^2k^2}{m}-E \\
\end{pmatrix}\begin{pmatrix}c_1\\c_2\\c_3 \end{pmatrix}=0,
\end{align}
Diagonalization of Eq. (\ref{eqevp}) yields the eigenvalues and eigenvectors.
\begin{align}
&\begin{pmatrix}c_1\\c_2\\c_3 \end{pmatrix}=\begin{pmatrix}\sqrt{\frac{1}{3}}\\0\\ \sqrt{\frac{2}{3}}\end{pmatrix}\,\,\text{for } E=\frac{\hbar^2k^2}{m},\\
&\begin{pmatrix}c_1\\c_2\\c_3 \end{pmatrix}=\begin{pmatrix}\sqrt{\frac{1}{3}}\\i\sqrt{\frac{1}{2}}\\ -\sqrt{\frac{1}{6}}\end{pmatrix}\,\,\text{for } E=\frac{\hbar^2k^2}{m}-\frac{2\hbar^2 \lambda k}{m},
\end{align}
\begin{equation}
\begin{pmatrix}c_1\\c_2\\c_3 \end{pmatrix}=\begin{pmatrix}\sqrt{\frac{1}{3}}\\-i\sqrt{\frac{1}{2}}\\ -\sqrt{\frac{1}{6}}\end{pmatrix}\,\,\text{for } E=\frac{\hbar^2k^2}{m}+\frac{2\hbar^2\lambda k}{m}.
\end{equation}
The eigenstates are orthonormal. Moreover, the same energy dispersion relations between the relative energy $E$ and the relative momentum $k$ are obtained by directly diagonalizing the non-interacting Hamiltonian in momentum space.
After writing the incident energy in the notation $E\equiv \hbar^2k_0^2/m$, the canonical momenta for channel 1 to 3 are found to be $k_1=k_0, k_2=\lambda+\sqrt{\lambda^2+k_0^2},$ and $k_3=-\lambda+\sqrt{\lambda^2+k_0^2}$. 
The set of regular solutions are
\begin{align}
\label{eqR0}
\small
\frac{\underline{f}(r)}{r}=
\begin{pmatrix}
\frac{N_1}{\sqrt{3}}k_1 j_0(k_1 r)&\frac{N_2}{\sqrt{{3}}}k_2 j_0(k_2 r)&\frac{N_3}{\sqrt{3}}k_3 j_0(k_3 r)\\
0&i \frac{N_2}{\sqrt{2}}k_2 j_1(k_2 r)&-i\frac{N_3}{\sqrt{2}}k_3 j_1(k_3 r)\\ 
\frac{N_1\sqrt{2}}{\sqrt{3}}k_1 j_2(k_1 r)&-\frac{N_2}{\sqrt{6}}k_2 j_2(k_2 r)&-\frac{N_3}{\sqrt{6}}k_3 j_2(k_3 r)
\end{pmatrix}.
\end{align}
The above solution can also be confirmed by projecting the plane wave solution in Eq.~(\ref{planewavesol}) onto the $\ket{J=0}$ subspace.
The normalization factors $\{N_1,N_2,N_3\}=\sqrt{\frac{2\mu}{\pi\hbar^2}}\{\sqrt{\frac{1}{k_1}},\sqrt{\frac{1}{k_2-\lambda}},\sqrt{\frac{1}{k_3+\lambda}}\}$ to each independent solution are added to ensure that their Wronskians with the irregular solutions (see appendix A) are identical, which in turn guarantees that the computed interaction $K$-matrix will be symmetric. This step is in fact equivalent to enforcing energy normalization of wave function in the case without spin-orbit coupling. 

The multichannel scattering formalism presented here is different from previous treatments when there is no single-particle potential existing even at large distances. In previous studies, one often chooses the asymptotically free states as the base pair of independent solutions to define phaseshifts or reaction matrices  and then study how short range interaction mixes different channels and causes particles to be scattered among those channels prior to being detected at large distances. And the incoming basis states expanded in the usual formulations of scattering theory having no long range channel coupling are diagonal solution matrices, which is not the case here as in Eq.~(\ref{eqR0}).  

After plugging in the free Green's matrix $\underline{G}(r,r')=\underline{\mathscr{G}}(r,r')/(rr')$ from Eq. (\ref{eqGred}) and the free radial wave function $\underline{R}_0(r)=\underline{f}(r)/r$ from Eq.~(\ref{eqR0}), we obtain the scattering solutions. The reaction matrix $K$ is determined through the correct asymptotic solution: 
\begin{equation}
\label{scattWF}
\underline{R}(r)|_{r\rightarrow\infty}\sim\frac{\underline{f}(r)}{r}-\frac{\underline{g}(r)}{r}\underline{K},
\end{equation}
where we find
\begin{equation}
\label{eqKM}
\underline{K}=\frac{-2a_s}{3}
\begin{pmatrix}
\frac{k_1}{2}&k_2\sqrt{\frac{k_1}{2(k_2+k_3)}}&k_3\sqrt{\frac{k_1}{2(k_2+k_3)}}\\
k_2\sqrt{\frac{k_1}{2(k_2+k_3)}}&\frac{k_2^2}{(k_2+k_3)}&\frac{k_2k_3}{(k_2+k_3)}\\
k_3\sqrt{\frac{k_1}{2(k_2+k_3)}}&\frac{k_2 k_3}{(k_2+k_3)}&\frac{k_3^2}{(k_2+k_3)}
\end{pmatrix}.
\end{equation}
From the $\underline{K}$ matrix, the $\underline{S}$ matrix is determined by the usual relation, $\underline{S}=(\underline{I}+i\underline{K})(\underline{I}-i\underline{K})^{-1}$. The unitarity of the $\underline{S}$ matrix is guaranteed by the real and symmetric reaction $K$ matrix 
as it is in Eq.~(\ref{eqKM}).

The scattered solutions defining the $\underline{S}$ matrix can be expressed as in Eq.~(\ref{scattWF}), 
{\small
\begin{align}
\label{pppp}
&R_{\ket{00,\vec{k}_1}\rightarrow\ket{00,\hat{r}}}\underset{r\rightarrow\infty}{\longrightarrow} 
\sqrt{2}\frac{S_{11}-1}{2ik_1}\frac{e^{ik_1r}}{r}\ket{00,\hat{r}}\\
&R_{\ket{00,\vec{k}_1}\rightarrow\ket{--,\hat{r}}}\underset{r\rightarrow\infty}{\longrightarrow} \sqrt{2}\sqrt{\frac{k_1}{k_2-\lambda}}\frac{S_{12}}{2ik_1}\frac{e^{ik_2r}}{r}\ket{--,\hat{r}}\\
&R_{\ket{00,\vec{k}_1}\rightarrow\ket{++,\hat{r}}}\underset{r\rightarrow\infty}{\longrightarrow} \sqrt{2}\sqrt{\frac{k_1}{k_3+\lambda}}\frac{S_{13}}{2ik_1}\frac{e^{ik_3 r}}{r}\ket{++,\hat{r}}\\
&R_{\ket{--,\vec{k}_2}\rightarrow\ket{00,\hat{r}}}\underset{r\rightarrow\infty}{\longrightarrow} 
\sqrt{2}\sqrt{\frac{(k_2-\lambda)}{k_1}}\frac{S_{21}}{2ik_2}\frac{e^{ik_1r}}{r}\ket{00,\hat{r}}\\
&R_{\ket{--,\vec{k}_2}\rightarrow\ket{--,\hat{r}}}\underset{r\rightarrow\infty}{\longrightarrow} \sqrt{2}\frac{S_{22}-1}{2ik_2}\frac{e^{ik_2r}}{r}\ket{--,\hat{r}}\\
&R_{\ket{--,\vec{k}_2}\rightarrow\ket{++,\hat{r}}}\underset{r\rightarrow\infty}{\longrightarrow}\sqrt{2} \sqrt{\frac{k_2-\lambda}{k_3+\lambda}}\frac{S_{23}}{2ik_2}\frac{e^{ik_3 r}}{r}\ket{++,\hat{r}}\\
&R_{\ket{++,\vec{k}_3}\rightarrow\ket{00,\hat{r}}}\underset{r\rightarrow\infty}{\longrightarrow} 
\sqrt{2}\sqrt{\frac{k_3+\lambda}{k_1}}\frac{S_{31}}{2ik_3}\frac{e^{ik_1r}}{r}\ket{00,\hat{r}}\\
&R_{\ket{++,\vec{k}_3}\rightarrow\ket{--,\hat{r}}}\underset{r\rightarrow\infty}{\longrightarrow} \sqrt{2}\sqrt{\frac{k_3+\lambda}{k_2-\lambda}}\frac{S_{32}}{2ik_3}\frac{e^{ik_2r}}{r}\ket{--,\hat{r}}\\
\label{nnnn}
&R_{\ket{++,\vec{k}_3}\rightarrow\ket{++,\hat{r}}}\underset{r\rightarrow\infty}{\longrightarrow} \sqrt{2}\frac{S_{33}-1}{2ik_3}\frac{e^{ik_3 r}}{r}\ket{++,\hat{r}}
\end{align}
}where the basis $\ket{\zeta\xi,\hat{r}}\equiv \ket{\zeta,\hat{r}}\ket{\xi,-\hat{r}}$.
From Eq.~(\ref{planewavesol}) and Eq.~(\ref{pppp})$\sim$(\ref{nnnn}), the incoming and outgoing current fluxes are determined by the velocity operator, $\vec{v}=\vec{p}/\mu+\hbar\lambda (\vec{s_1}-\vec{s}_2)/m$. The integrated partial cross sections are found by integrating the flux ratio over all solid angles. The total cross section for particles incident in channel $\alpha$ with some helicity to be scattered into channel $\beta$ of another helicity is 
\begin{equation}
\sigma_{\alpha\beta}=\frac{2\pi}{k_\alpha^2}|S_{\alpha\beta}-\delta_{\alpha\beta}|^2,
\end{equation}
where $k_\alpha$ is the canonical momentum in the incoming state. 
Using the SOC strength as the unit of the momentum, the cross section can be rescaled as a function of the dimensionless quantity, $\lambda a_s$, by choosing the unit of cross section as $1/\lambda^2$. From the estimation in \cite{Ueda}\cite{pro_2}, a realistic achievable value of the SOC strength is around $1\sim10/\mu m$, and this guides our chosen values of $\lambda a_s$ in the plots shown for the cross sections.  
\begin{figure}
\begin{center}
\includegraphics[width=0.4\textwidth]{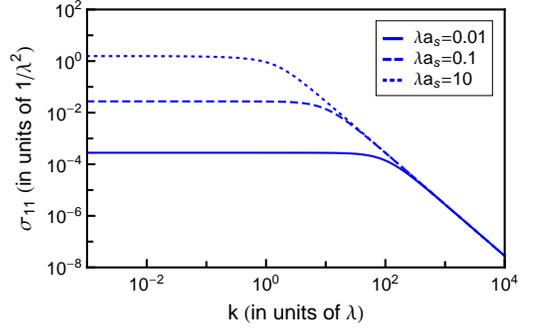}
\caption{The cross sections from the incoming state $\ket{00,\vec{k}_1}$ to the outgoing state $\ket{00,\hat{r}}$ for different values of $\lambda a_s$.}
\label{sigma11}
\end{center}
\end{figure}
\begin{figure}
\begin{center}
\includegraphics[width=0.4\textwidth]{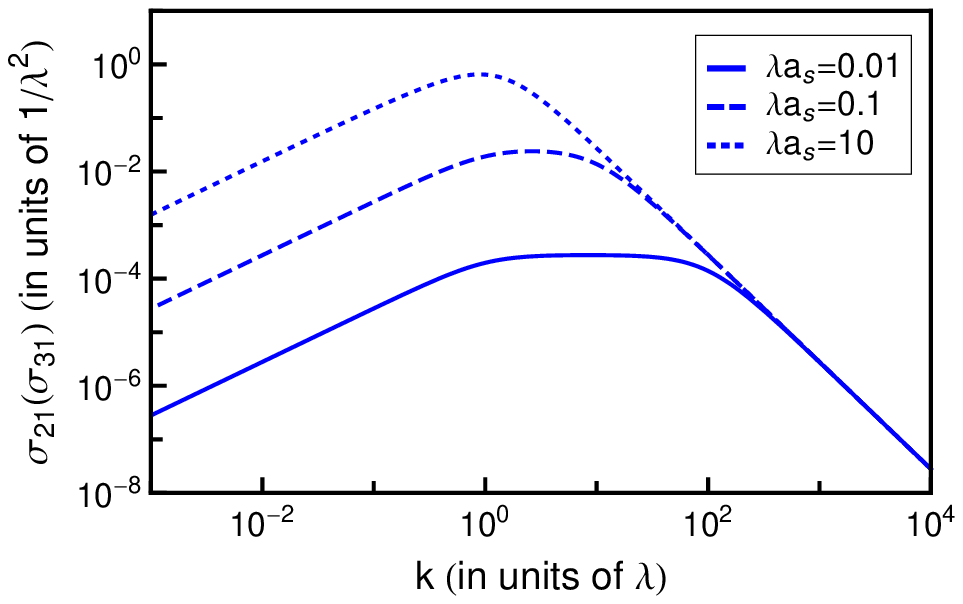}
\caption{The cross sections from the incoming state $\ket{--,\vec{k}_2}$ ($\ket{++,\vec{k}_3}$) to the outgoing state $\ket{00,\hat{r}}$ for different values of $\lambda a_s$.}
\label{sigma21}
\end{center}
\end{figure}
\begin{figure}
\begin{center}
\includegraphics[width=0.4\textwidth]{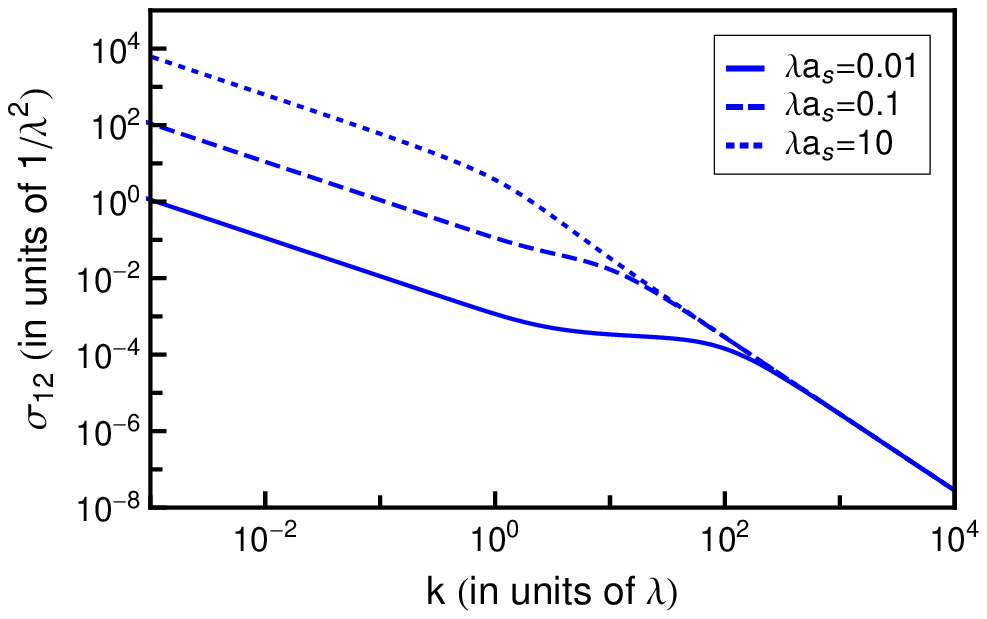}
\caption{The cross sections from the incoming state $\ket{00,\vec{k}_1}$ to the outgoing state $\ket{--,\hat{r}}$ for different values of $\lambda a_s$.}
\label{sigma12}
\end{center}
\end{figure}
\begin{figure}
\begin{center}
\includegraphics[width=0.4\textwidth]{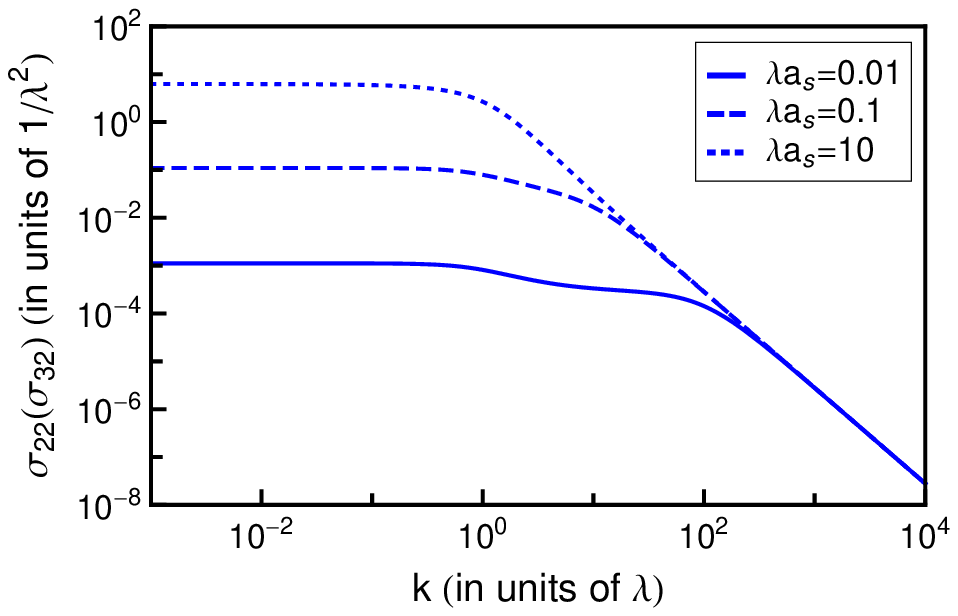}
\caption{The cross sections from the incoming state $\ket{--,\vec{k}_2}$ ($\ket{++,\vec{k}_3}$) to the outgoing state $\ket{--,\hat{r}}$ for different values of $\lambda a_s$.}
\label{sigma22}
\end{center}
\end{figure}
\begin{figure}
\begin{center}
\includegraphics[width=0.4\textwidth]{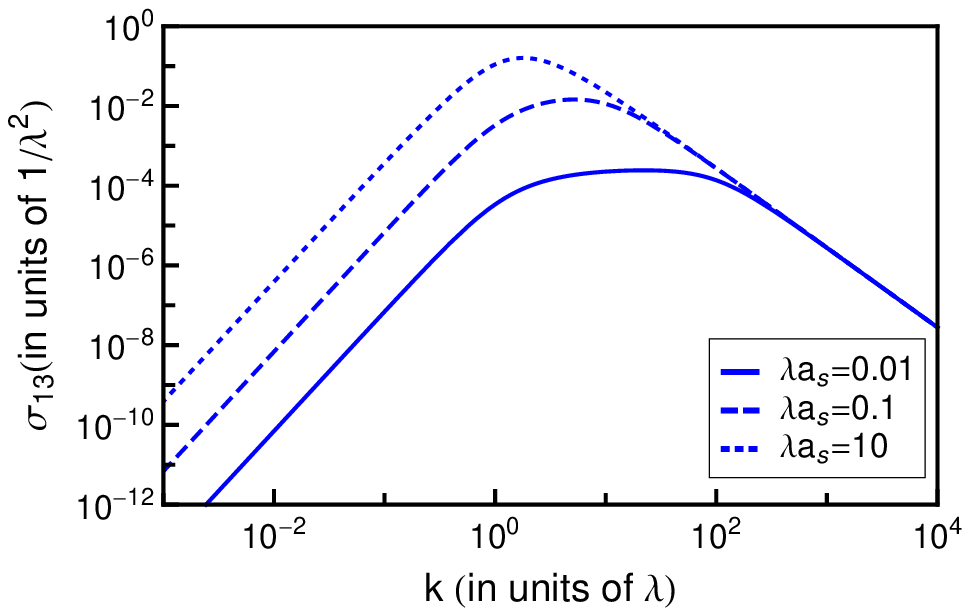}
\caption{The cross sections from the incoming state $\ket{00,\vec{k}_1}$ to the outgoing state $\ket{++,\hat{r}}$ for different values of $\lambda a_s$.}
\label{sigma13}
\end{center}
\end{figure}
\begin{figure}
\begin{center}
\includegraphics[width=0.4\textwidth]{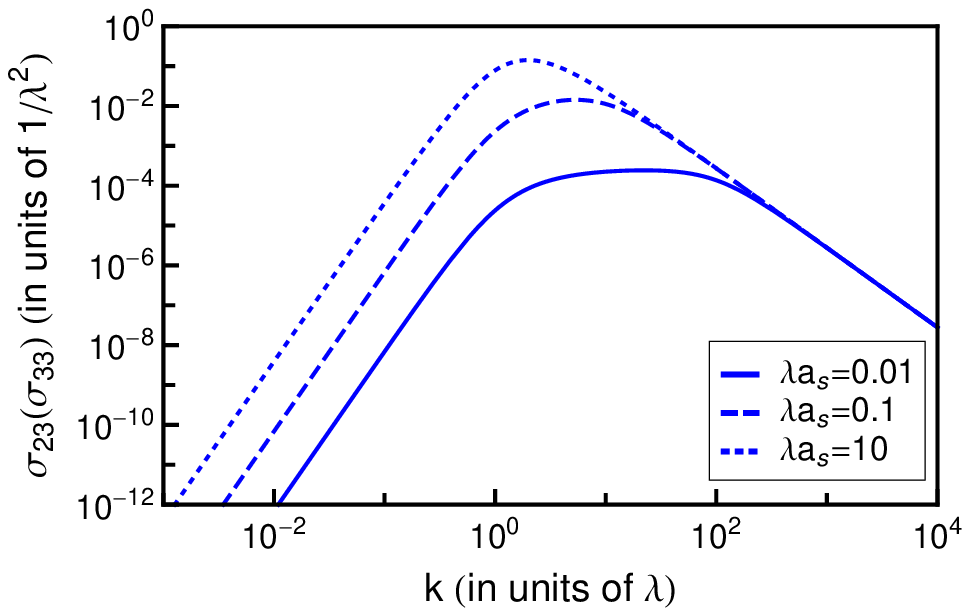}
\caption{The cross sections from the incoming state $\ket{--,\vec{k}_2}$ ($\ket{++,\vec{k}_3}$) to the outgoing state $\ket{++,\hat{r}}$ for different values of $\lambda a_s$.}
\label{sigma23}
\end{center}
\end{figure}

Turning off the spin-orbit coupling, all of the cross sections return to the classic scaling behaviors: insensitive to energy in the low $k$ limit and proportional to $k^{-2}$ in the higher $k$ limit. This transition happens when $k\sim \frac{1}{a_s}$. The unusual factor, $8/9$, is due to the choice of the helicity basis.
\begin{equation}
\sigma_{ij}=\frac{8\pi a_s^2}{9+9a_s^2k^2}
\approx\begin{cases} \frac{8\pi a_s^2}{9} & k\ll\frac{1}{a_s} \\ \frac{8\pi}{9k^2} & k\gg \frac{1}{a_s} \end{cases}\; \; \forall \;i,j.
\end{equation}
In the limit of high $k$ (but still low energy) scattering, the scattering cross section becomes insensitive to the existence of spin-orbit coupling. This is expected since at small distance, the short-range interaction dominates and the physics of SOC becomes insignificant. All cross sections are combined into the same curve in this limit, namely $\sigma_{ij}\sim 8\pi/(9k^2)$ as in the high energy limit of non-SOC cases. 

The effect of spin-orbit coupling becomes important as energy decreases below the energy scale set by SOC. This explains why there exists a transitional plateau when $\lambda<k<1/a_s$. This is of course possible only when the interatomic interaction is weaker than SOC.
The cross sections in the low $k$ limit are no longer energy independent and show some unusual features. The cross sections in different channels are characterized by different power laws at very low temperatures. Scattering is enhanced or suppressed depending on which outgoing channels are taken. The scaling laws are summarized as follows.
\begin{align}
&\sigma_{11}\approx \frac{8\pi a_s^2}{9+16(\lambda a_s)^2} &\text{ for } k<\lambda\\
&\sigma_{21}\approx \frac{8\pi a_s^2}{9+16(\lambda a_s)^2}\frac{k}{\lambda} &\text{ for } k<\lambda\\
&\sigma_{12}\approx \frac{32\pi a_s^2}{9+16(\lambda a_s)^2} \bigg(\frac{k}{\lambda}\bigg)^{-1}  &\text{ for } k<\lambda\\
&\sigma_{22}\approx \frac{32\pi a_s^2}{9+16(\lambda a_s)^2}  &\text{ for } k<\lambda\\
&\sigma_{13}\approx \frac{2\pi a_s^2}{9+16(\lambda a_s)^2}\bigg(\frac{k}{\lambda}\bigg)^3  &\text{ for } k<\lambda\\
&\sigma_{23}\approx \frac{2\pi a_s^2}{9+16(\lambda a_s)^2}\bigg(\frac{k}{\lambda}\bigg)^4  &\text{ for } k<\lambda
\end{align}

Even when the cross sections in some channels ($\sigma_{11}$, $\sigma_{22}$, and $\sigma_{32}$) at low temperatures in the presence of SOC are insensitive to energy, the effect of SOC can still be seen by studying the threshold values. When $\lambda a_s \lesssim~1$ ($\lambda a_s \gtrsim~1$), the cross section $\sigma_{22}$ or $\sigma_{32}$ is increased (decreased) from the non-SOC case. For the particular channel in $\sigma_{11}$, the cross section is smaller than the non-SOC case until $\lambda a_s$ reaches 1 from above. Therefore, the effect of SOC cannot be differentiated even in the very low energy limit when $\lambda a_s\lesssim 1$ in the $\ket{00,\hat{r}}\rightarrow \ket{00,\hat{r}}$ channel.
  
\begin{figure}
\begin{center}
\includegraphics[width=0.44\textwidth]{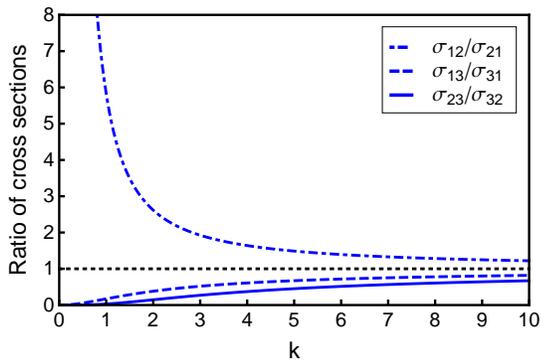}
\caption{(Color online) The ratio of cross sections. The lowest helicity state, $k_2$ dominates. For cases without spin-orbit coupling, the ratio is equal to unity, which is drawn as a black dotted line for reference.}
\label{ratio}
\end{center}
\end{figure}

From Fig.~\ref{sigma11} to Fig.~\ref{sigma23}, processes where particles transfer to the lowest helicity state labeled by $k_2$ are enhanced compared to the non-SOC case. Moreover, particles are preferentially scattered into the $k_2$ channel where the particle's momentum is antiparallel to its spin direction, regardless of their incidence channel. Fig. (\ref{ratio}) shows that the $k_2$ channel will dominate among all helicity states, as can be seen by comparing the ratios of the different scattering cross sections.
\begin{equation}
\frac{\sigma_{\alpha\beta}}{\sigma_{\beta\alpha}}=\bigg(\frac{k_\beta}{k_\alpha}\bigg)^2. 
\end{equation}
We think the SOC system, which can also be interpreted as spins in a momentum-dependent B field, is an analog to an antiferromagnetic system. The magnetic potential energy is minimized when spin is antiparallel to the direction of field. Similarly, particles would like to stay in their lowest helicity states when the particle's spin has a reverse direction to its momentum. The spontaneous handedness appears in parity-breaking systems when interaction can cause fluctuation among system's eigenstates. 

The bound state information can also be predicted by searching for the poles of $\underline{S}$ matrix. The scattering threshold energy here is $E_T=-\hbar^2\lambda^2/2\mu$. For energy $E<E_T$, all channels are closed. We take the following analytical continuation:
\begin{align}
&k_1=i \kappa\\
&k_2=i\sqrt{\kappa^2-\lambda^2}+\lambda\\
&k_3=i\sqrt{\kappa^2-\lambda^2}-\lambda,
\end{align}
where $\kappa$ is chosen to be positive so the exponentially growing part in the incoming scattering wave functions is killed. The bound state wave function can be found by plugging the continuation into the outgoing wave functions. The new feature brought into the bound state wave function by SOC is that the function is now decaying exponentially with an spatial oscillation whose frequency is set by $\lambda$.
The binding energy for the bound pair is given by $E_b=E_T+\hbar^2\kappa^2/2\mu>0$, where $\kappa$ is found by solving $\text{Det}(\underline{I}-i\underline{K})=0$.
The binding energy returns to the usual case with an overall constant shift, depending on the strength of SOC, when $a_s$ is approaching zero from the positive side.
\begin{align}
E_b=
\begin{cases}
\frac{\hbar^2}{2\mu a_s^2}+\frac{\hbar^2\lambda^2}{2\mu}&\frac{1}{\lambda a_s}\rightarrow +\infty\\
\frac{(9-\sqrt{33})\hbar^2\lambda^2}{12\mu}+4\sqrt{\frac{2}{11}-\frac{1}{\sqrt{33}}}\frac{\hbar^2\lambda}{\mu a_s}  &\frac{1}{\lambda a_s}\rightarrow 0\\
\frac{2\hbar^2\lambda^4 a_s^2}{9\mu}  &\frac{1}{\lambda a_s}\rightarrow -\infty.
\end{cases}
\end{align}
One interesting effect from SOC shows up in the small and negative $a_s$ limit. The binding energy scales algebraically as $\lambda^4 a_s^2$, which indicates that the existence of a two-body bound state no matter how small and attractive the scattering length is as long as SOC exists \cite{bound,pro_1}. The two-body bound state information paves the way to more complex trimer systems.
\begin{figure}[!htb]
        \centering
        \includegraphics[width=0.44\textwidth]{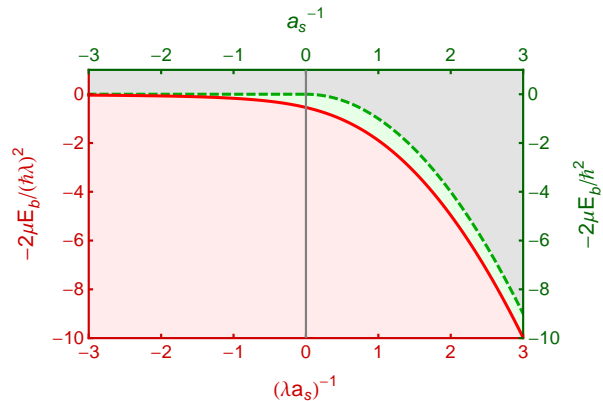}
    \caption{(Color online) The two-body binding energy for two spin-1 bosons in the presence (red solid line) and absence (green dashed line) of SOC. The bottom and left axes apply to the red curve, and the top and right axes apply to the green dashed curve.}
\end{figure}

%%%%%%%%%%%%%%%%%%%%%%%%%%%%%%%%%%%%%%%%%%%%%%%%%%%%%%%%%%%%%%%%%%%%%%%%
% conclusions
%%%%%%%%%%%%%%%%%%%%%%%%%%%%%%%%%%%%%%%%%%%%%%%%%%%%%%%%%%%%%%%%%%%%%%%%
\section{Conclusions}
To summarize, the present treatment extends the work of the previous studies by \cite{Duan} and develops a more general treatment of ultracold scattering in the presence of isotropic spin-orbit coupling.  Our formulation can apply to any two identical particles with arbitrary spin, in any total angular momentum subspace. This should enable a deeper understanding of low energy scattering (or two-body bound states) in the presence of an artificial gauge field, especially those which are non-abelian. The non-abelian gauge fields become possible when atoms' internal degrees of freedom are utilized.

\section{Acknowledgement}
We would like to thank Panagiotis Giannakeas for his helpful discussions and Baochun Yang and Jes\'us P\'erez-R\'ios for their suggestions while preparing for this manuscript. This work was supported by NSF under grant number PHY-130690, and by a Purdue University Research Incentive Grant from the Office of the Vice President for Research. 

%%%%%%%%%%%%%%%%%%%%%%%%%%%%%%%%%%%%%%%%%%%%%%%%%%%%%%%%%%%%%%%%%%%%%%%%
% appendix A
%%%%%%%%%%%%%%%%%%%%%%%%%%%%%%%%%%%%%%%%%%%%%%%%%%%%%%%%%%%%%%%%%%%%%%%%
\appendix
\section{Derivation of Green's matrix}
The reduced radial Green's matrix satisfies the following coupled differential equation:
\begin{equation}
\label{a1}
\bigg(-\frac{\hbar^2}{2\mu}\frac{d^2}{dr^2}\delta_{ij}+iA_{ij}\frac{d}{dr}+B_{ij}\bigg)\mathscr{G}_{jk}(r,r')=-{\delta_{ik}}{\delta(r-r')},
\end{equation}
where $A$ is a real and symmetric matrix and $B$ is a hermitian matrix without involving any derivative. The index $\{i,j,k\}$ run from $1$ to $n$. Summation over $j$ is implied. Although we study this particular type of coupled equations in Eq.~(\ref{a1}), the procedures provided below is general and be applied to any type of coupled equations. The Green's matrix is used to emphasize the nature of {\em coupled} differential equations. If there is only one equation, the Green's matrix has only one component, so returns to the commonly termed Green's function. 

The Green's matrix is constructed with the assistance of $n$ regular and $n$ irregular solutions of the homogenous equations,
\begin{align}
\label{a2}
&\bigg(-\frac{\hbar^2}{2\mu}\frac{d^2}{dr^2}\delta_{ij}+iA_{ij}\frac{d}{dr}+B_{ij}\bigg) f_{j\alpha}(r)=0  \\
\label{a3}
&\bigg(-\frac{\hbar^2}{2\mu}\frac{d^2}{dr^2}\delta_{ij}+iA_{ij}\frac{d}{dr}+B_{ij}\bigg) g_{j\alpha}(r)=0.
\end{align}
Each column of $\underline{f}$ and $\underline{g}$ correspond to one independent regular/irregular solutions. For convenience, the notation $\underline{f}_\alpha$ and $\underline{g}_\alpha$ $(\alpha=1,2,...,n)$ for each independent regular and irregular solution will be used. The regular solution has to satisfy 
\begin{equation}
\underline{f}_\alpha(r=0)=\underline{0}.
\end{equation}
The boundary condition for the irregular function is satisfied by requiring a $\pi/2$ phase lag to the regular solution at very large distance, $r\rightarrow\infty$.

Knowing that the regular and irregular solutions of the homogenous differential equation, we make the ansatz for the reduced Green's matrix: 
\begin{eqnarray}
\label{ansatzG}
{\underline{\mathscr{G}}}(r,r')=\begin{cases}
    \underline{f}(r)\underline{S}(r')  & \text{for\,\,\,\,\,} r<r' , \\
     \underline{g}(r)\underline{T}(r') & \text{for\,\,\,\,\,} r>r'.
\end{cases}
\end{eqnarray}
The next step is to match the expressions for the reduced Green's matrices at $r=r'$ and to apply the appropriate derivative discontinuity, 
\begin{eqnarray}
\label{bcG}
&&\underline{f}(r')\underline{S}(r')=\underline{g}(r')\underline{T}(r')\\
\label{bcG2}
&&\lim_{\epsilon \rightarrow0}\frac{d}{dr}\underline{\mathscr{G}}(r,r')|_{r'-\epsilon}^{r'+\epsilon}=\frac{2\mu}{\hbar^2}\underline{I}.
\end{eqnarray}
From Eq.~(\ref{bcG}), the matrix $\underline{S}(r')$ can be rewritten in terms of $\underline{T}(r')$ as
\begin{equation}
\label{ST}
\underline{S}(r')=\underline{f}^{-1}(r')\underline{g}(r')\underline{T}(r').
\end{equation}
Application of Eq.~(\ref{ST}) to Eq.~(\ref{ansatzG}) reduces Eq.~(\ref{bcG2}) into an algebraic equation for the matrix $\underline{T}(r')$,
\begin{align}
\bigg(\frac{d\underline{g}(r')}{dr'}-\frac{d\underline{f}(r')}{dr'}\underline{f}^{-1}(r')\underline{g}(r')\bigg)\underline{T}(r')=\frac{2\mu}{\hbar^2}\underline{I}
\end{align}
Therefore, 
\begin{equation}
\label{TT}
\underline{T}(r')=\frac{2\mu}{\hbar^2}[\underline{g}'(r')-\underline{f}'(r')\underline{f}^{-1}(r')\underline{g}(r')]^{-1}.
\end{equation}
Combining Eq. (\ref{ST}) and (\ref{TT}), the matrices $\underline{S}$ and $\underline{T}$ are found to be 
\begin{align}
\label{Smat}
&\underline{S}=\frac{2\mu}{\hbar^2}\times \underline{f}^{-1}(\underline{g}'\underline{g}^{-1}-\underline{f}'\underline{f}^{-1})^{-1}\\
\label{Tmat}
&\underline{T}=\frac{2\mu}{\hbar^2}\times \underline{g}^{-1}(\underline{g}'\underline{g}^{-1}-\underline{f}'\underline{f}^{-1})^{-1}.
\end{align}
It can be shown further that the Green's matrix in Eq.~(\ref{ansatzG}) with Eq.~(\ref{Smat}) and Eq.~(\ref{Tmat}) indeed returns to the familiar form.
\begin{eqnarray}
\label{finalG}
{\underline{\mathscr{G}}}(r,r')=\begin{cases}
   \pi \underline{f}(r)\underline{g}^\dagger(r') & \text{for\,\,\,\,\,} r<r' , \\
   \pi \underline{g}(r)\underline{f}^{\dagger}(r')& \text{for\,\,\,\,\,} r>r'.
\end{cases}
\end{eqnarray}
Before we do that, we need first to prove that the analogues of the Wronskian for Eq.~(\ref{a1}) are 
\begin{align}
\label{eqW1}
&\frac{\hbar^2}{2\mu}({\underline{g}}^{\prime\dagger} \underline{f}-\underline{g}^\dagger \underline{f}^\prime)+i\underline{g}^\dagger \underline{A} \underline{f}=\underline{C}\\
\label{eqW2}
&\frac{\hbar^2}{2\mu}({\underline{f}}^{\prime\dagger} \underline{f}-\underline{f}^\dagger \underline{f}^\prime)+i\underline{f}^\dagger \underline{A} \underline{f}=\underline{0}\\
\label{eqW3}
&\frac{\hbar^2}{2\mu}({\underline{g}}^{\prime\dagger} \underline{g}-\underline{g}^\dagger \underline{g}^\prime)+i\underline{g}^\dagger \underline{A} \underline{g}=\underline{0},
\end{align}
where $\underline{C}$ is a $r$-independent constant matrix and will be determined later by the requirement of energy normalization, and $\underline{0}$ is a zero matrix.
The above set of Wronskians is shown below. 
Application of ${\underline{g}}_\beta^{\dagger}$ to Eq.~(\ref{a2}) and $\underline{f}_\alpha^\dagger$ to Eq.~(\ref{a3}) separately yields
\begin{align}
\label{y1C}
&-\underline{g}_\beta^\dagger \frac{\hbar^2}{2\mu}\underline{f}''_\alpha+i\underline{g}_\beta^\dagger\underline{A}\underline{f}'_\alpha+\underline{g}_\beta^\dagger \underline{B}{\underline{f}_\alpha}=0\\
\label{y2C}
&-\underline{f}_\alpha^\dagger \frac{\hbar^2}{2\mu} \underline{g''_\beta}+i\underline{f}_\alpha^\dagger\underline{A}{\underline{g}'_\beta}+\underline{f}_\alpha^\dagger \underline{B}{\underline{g}_\beta}=0,
\end{align}
Subtracting the complex conjugate of Eq. (\ref{y2C}) from Eq. (\ref{y1C}) gives the following equality,
\begin{align}
\label{diff}
\sum_j \frac{-\hbar^2}{2\mu}(g_{j\beta}^*f_{j\alpha}''-f_{j\alpha}{g_{j\beta}^{*\prime\prime}})+i\sum_{j,k}(f_{j\alpha}'g_{k\beta}^*+f_{j\alpha}{g_{k\beta}^{*\prime}})A_{jk}=0.
\end{align}
All the matrices are expressed in terms of their matrix elements. 
The properties of the matrices $\underline{A}$ and $\underline{B}$ are used to derive the above identity. After integration of both sides of Eq.~(\ref{diff}) over $r$ from $a$ to $b$, one has 
\begin{align}
\sum_j\frac{\hbar^2}{2\mu}(f_{j\alpha}{g_{j\beta}^{*\prime}}- f_{j\alpha}' g_{j\beta}^*)\big|_a^b+i\sum_{j,k} f_{j\alpha} A_{jk} g_{k\beta}^* \big|_a^b=0.
\end{align}
The above formula is of course true for any range $[a,b]$, so we know that in matrix notation the following expression should be a constant, which is position-independent.
\begin{equation}
\frac{\hbar^2}{2\mu}({{\underline{g}}_\beta^{\prime\dagger}} \underline{f}_\alpha-\underline{g}_\beta^\dagger \underline{f}_\alpha')+i\underline{g}_\beta^\dagger \underline{A} \underline{f}_\alpha=c\delta_{\alpha\beta}.
\end{equation}
Combining all the independent regular and irregular solutions, the ``modified'' Wronskian in Eq.~(\ref{eqW1}) is derived.
The other two Wronskians in Eq.~(\ref{eqW2}) and Eq.~(\ref{eqW3}) can be proved in a similar way. Noticing that the matrix $\underline{A}$ is proportional to the spin-orbit coupling strength, removal of the second term on the left hand side of Eq.~(\ref{eqW1})$\sim$(\ref{eqW3}) reduce to the familiar formula. The constant matrix $\underline{C}$ is determined by energy normalization. Application of energy normalization is important to guarantee unitarity of the scattering $\underline{S}$ matrix, which reflects flux conservation. The proper energy normalization gives
$\underline{C}=\frac{1}{\pi}\underline{I}$.

The missing piece connecting Eq.~(\ref{Smat}) and Eq.~(\ref{Tmat}) to Eq.~(\ref{finalG}) can be put together now. Taking the conjugate transpose of Eq.~(\ref{eqW1}), we find
\begin{align}
\label{midstep}
\frac{\hbar^2}{2\mu}(\underline{f}^{\dagger}\underline{g}^\prime-\underline{f}^{\prime\dagger} \underline{g})-i\underline{f}^\dagger \underline{A} \underline{g}=\frac{1}{\pi}\underline{I}
\end{align}
Applying $(\underline{f}^{\dagger})^{-1}$ to the left-hand side of Eq.~(\ref{midstep}) and $\underline{g}^{-1}$ to the right-hand side reduces the above equation into the following
\begin{align}
\label{ggff}
\frac{\hbar^2}{2\mu}\underline{g}^{\prime}\underline{g}^{-1}-(\frac{\hbar^2}{2\mu}\underline{f}^\prime \underline{f}^{-1}-i \underline{A})^\dagger=\frac{1}{\pi}(\underline{f}^\dagger)^{-1}\underline{g}^{-1}.
\end{align}
Also from Eq.~(\ref{eqW2}) the relation can be derived,
\begin{align}
\frac{\hbar^2}{2\mu}\underline{f}^\prime \underline{f}^{-1}=(\frac{\hbar^2}{2\mu}\underline{f}^\prime \underline{f}^{-1}-i\underline{A})^\dagger.
\end{align}
Therefore, Eq.~(\ref{ggff}) is further simplified to be
\begin{align}
\label{Ieq}
\underline{g}^{\prime}\underline{g}^{-1}-\underline{f}^\prime \underline{f}^{-1}=\frac{2\mu}{\hbar^2\pi}({\underline{f}^\dagger})^{-1}\underline{g}^{-1}.
\end{align}
Plugging Eq.~(\ref{Ieq}) into Eq.~(\ref{Tmat}), it is straightforward to see that $\underline{T}=\pi f^\dagger$. Similarly, the matrix $\underline{S}$ is proved to be $\underline{S}=\pi\underline{g}^\dagger$.

\bibliography{ref}

\end{document}